\documentclass[structabstract]{aa}  
\usepackage{graphicx}
\usepackage{txfonts}
\usepackage{natbib}

\usepackage{longtable}
\usepackage{comment}

\begin{document}
   \title{Evolution of the central stars of young planetary nebulae}

   \author{M. Hajduk\inst{1}, P. A. M. van Hoof\inst{2}, A. A. Zijlstra\inst{3}
          }

   \institute{Nicolaus Copernicus Astronomical Center, 
ul. Rabia\'{n}ska 8, 87-100 Toru\'{n}, Poland \and
    Royal Observatory of Belgium, Ringlaan 3, 1180 Brussels, Belgium \and
    Jodrell Bank Centre for Astrophysics, Alan Turing Building,
Manchester M13 9PL, UK}


  \abstract{}
{The evolution of central stars of planetary nebulae was so far documented in just a few cases. However, spectra collected a few decades ago may provide a good reference for studying the evolution of central stars using the emission line fluxes of their nebulae. We investigated evolutionary changes of the [O\,{\sc iii}] 5007\,\AA\ line flux in the spectra of planetary nebulae.}
{We compared nebular fluxes collected during a decade or longer. We used literature data and newly obtained spectra. A grid of Cloudy models was computed using existing evolutionary models, and the models were compared with the observations.}
{An increase of the [O\,{\sc iii}] 5007\,\AA\ line flux is frequently observed in young planetary nebulae hosting H-rich central stars. The increasing nebular excitation is the response to the increasing temperature and hardening radiation of the central stars. We did not observe any changes in the nebular fluxes in the planetary nebulae hosting late-type Wolf-Rayet (WR) central stars. This may indicate a slower temperature evolution (which may stem from a different evolutionary status) of late-[WR] stars.}
{In young planetary nebulae with H-rich central stars, the evolution can be followed using optical spectra collected during a decade or longer. The observed evolution of H-rich central stars is consistent with the predictions of the evolutionary models provided in the literature. Late-[WR] stars possibly follow a different evolutionary path.}

\keywords{interstellar medium: planetary nebulae: general -- stars: evolution -- stars: AGB and post-AGB}
\maketitle
%

\section{Introduction}

Low- and intermediate-mass stars with initial masses of $\rm 1-8 \, M_\odot$
spend most of their life on the main sequence, steadily burning hydrogen in
their centers. When hydrogen becomes exhausted in the center, the star leaves
the main sequence and evolves to the red giant branch. Hydrogen is being burned
in a shell around the helium core up to a moment when helium ignites in the
core. After helium is exhausted in the center, the nuclear burning continues in
the helium shell around the C/O degenerate core. The star ascends the asymptotic
giant branch (AGB). In the last phase of the AGB evolution, helium flashes
punctuate the periods of quiescent hydrogen shell burning. The stellar envelope
is extended and loosely bound. The mass-loss rivals the rate of nuclear burning.
As a result, the star ejects almost the entire envelope.

During the post-AGB phase, the star rapidly heats up and ionizes the lost mass,
which forms a planetary nebula (PN). The stellar temperature increase from
$<$10\,kK to $>$100\,kK within $10^3$--$10^4$\,yr. The pace of the evolution
depends critically on the final mass of the star \citep{1995A&A...299..755B}. As
the star heats up, the ionization fronts of $\rm H^0$, $\rm He^0$, and $\rm
He^+$ penetrate the ejected envelope, and the excitation level of the PN
increases. The star terminates its evolution as a white dwarf.

Observers rarely search for temporal variations in spectra of PNe. However, a
change of the nebular fluxes was observed in the optical, radio, or ultraviolet
in a few cases and was explained by the temperature evolution of their central
stars. 

\citet{1992PASP..104..339F} summarized earlier works on the variability of the
spectra of PNe IC\,4997 and NGC\,6572 and presented their the spectra taken with
the International Ultraviolet Explorer (IUE) and in the optical. They
interpreted their observations in terms of the increasing stellar temperatures.

\citet{1993IAUS..155..484H} reported the decline of the central star of
NGC\,2392 in the IUE spectra. \citet{1993AAS...18310104H} reported evolutionary
fading seen in the IUE observations of other O-type central stars.

\citet{2005A&AT...24..291K} spectrally monitored of six PNe for a few decades.
She reported significant changes in two low-excitation objects (M\,1-11 and
M\,1-6).

\citet{2008ApJ...681.1296Z} reported the radio flux evolution of the NGC\,7027,
a secondary flux calibrator. The flux evolution was modeled with the nebular
expansion and decrease of the ionizing photons, consistent with the evolutionary
models.

The flux evolution is clearly observed in the central stars of PNe that
experience a very late thermal pulse: V4334\,Sgr and V605\,Aql
\citep{2008ASPC..391..155V}.

Recently, \citet{2014arXiv1405.0800H} observed a significant flux change in the
spectra of the low-excitation PN Hen\,2-260 that hosts an O-type central star.
The comparison of the data obtained during 11 years indicated a 50\% increase of
the [O\,{\sc iii}] 5007\,\AA\ line flux and an increase of other high-ionization
lines. We interpreted our finding in terms of the expansion of the $\rm O^{+}$
ionization front as a response to the evolution of the central star. The derived
heating rate of the central star and the age of the nebula were consistent with
the post-AGB evolutionary tracks by \citet{1995A&A...299..755B}.

In this paper we compare the [O\,{\sc iii}] 5007\,\AA\ line fluxes of PNe
collected in different years and search for the evolutionary changes. We use the
fluxes published in the literature (Table~\ref{obs}). We also observed a sample
of PNe and added them to our analysis (Table~\ref{ratios}). The observed
evolution of the [O\,{\sc iii}] 5007\,\AA\ line fluxes is compared with the
evolution modeled using the evolutionary models and a photoionization code. We
also split the sample into different groups whose central stars showed different
surface abundances to constrain their evolutionary status.

\section{Data}

We used optical spectra available in the literature. The data set is not
uniform. The spectra were obtained using various telescopes and spectrograph
configurations. The year(s) of observations, apertures used, and the spectral
resolution for different data sets are shown in Table~\ref{obs}.

Very many objects were observed in these surveys. We selected a sample of about
50 objects that were observed by at least two different groups. We excluded most
of the objects observed by \citet{1996PASP..108..980K}, because their
observation dates were not precisely specified. We also excluded fluxes from
\citet{1994MNRAS.271..257K} from our analysis, because they were averaged from
spectra taken in different years and using different slit widths.

\citet{1996PASP..108..980K} and \citet{2004MNRAS.349.1291E} used circular
apertures. Most of spectra published by \citet{1994MNRAS.271..257K},
\citet{2003PASP..115...80K}, and \citet{2010ApJ...724..748H} were obtained at
the position angle (PA) of 90 degree. \citet{2005MNRAS.362..424W} observed most
of the objects at a PA of 0 degree.

\begin{table}
\caption{Spectroscopic surveys of PNe. \label{obs}}
\begin{center}
{\tiny
\begin{tabular}{cccccc}
\hline \hline
authors			&period		&aper. ['']&resol. [\AA]& $\chi ^2$\\
\hline
{\citet{1994MNRAS.271..257K}}	&1978-89	&	&	&1.60\\
{\citet{1996PASP..108..980K}}	&1981-1986	&8	&10	&\\
{\citet{1992secg.book.....A}}	&1983-89	&4x4, 4x7&10	&\\
{\citet{2004MNRAS.349.1291E}}	&1992		&6.4	&	&\\
{\citet{1996A&A...307..215C}}	&1992-93	&1.5x3.4&4	&0.73\\
{\citet{2003PASP..115...80K}}	&1996-97,1999	&5	&8.6	&\\
{\citet{2004A&A...414..211E}}	&2001-02	&2	&	&\\
{\citet{2009A&A...500.1089G}}	&2001-02	&4	&	&\\
{\citet{2005MNRAS.362..424W}}	&2001		&1	&4	&\\
{\citet{2010ApJ...724..748H}}	&2007		&2	&7	&0.95\\
{\citet{2012A&A...538A..54G}}	&2009-2010	&1	&0.14-0.17&1.12\\
\hline
\end{tabular}}
\end{center}
\end{table}

We supplemented the literature data with our observations of 20 PNe
(Table~\ref{ratios}). Low-resolution spectroscopy was performed with the 1.9m
Radcliffe and SALT telescope at the South African Astronomical Observatory
(SAAO) and the 0.9m Schmidt-Cassegrain telescope at the Toru\'{n} observatory.
The Radcliffe 1.9m telescope low-resolution spectra  were obtained using the
300\,l/mm grating. The SALT spectra were taken using the 900\,l/mm grating. We
used the 300\,l/mm grating on the 0.9m Toru\'{n} telescope.

The SALT spectrum obtained on 2012 May 7, all the Toru\'{n} spectra, and the
Radcliffe spectra taken on 2013 July 21 were obtained in photometric conditions.
All the remaining data were taken in nonphotometric conditions.

The PA was set to 0 degree in all the instruments. All the spectra were flux
calibrated using the observations of standard stars shown in Table~\ref{ratios}.
Standard stars were observed with the same slit width as the targets at
airmassess that exceeded 1.25 only in one case.

\section{Results, observational uncertainties, and biases}

  \begin{figure}
   \centering
   \includegraphics[width=8cm]{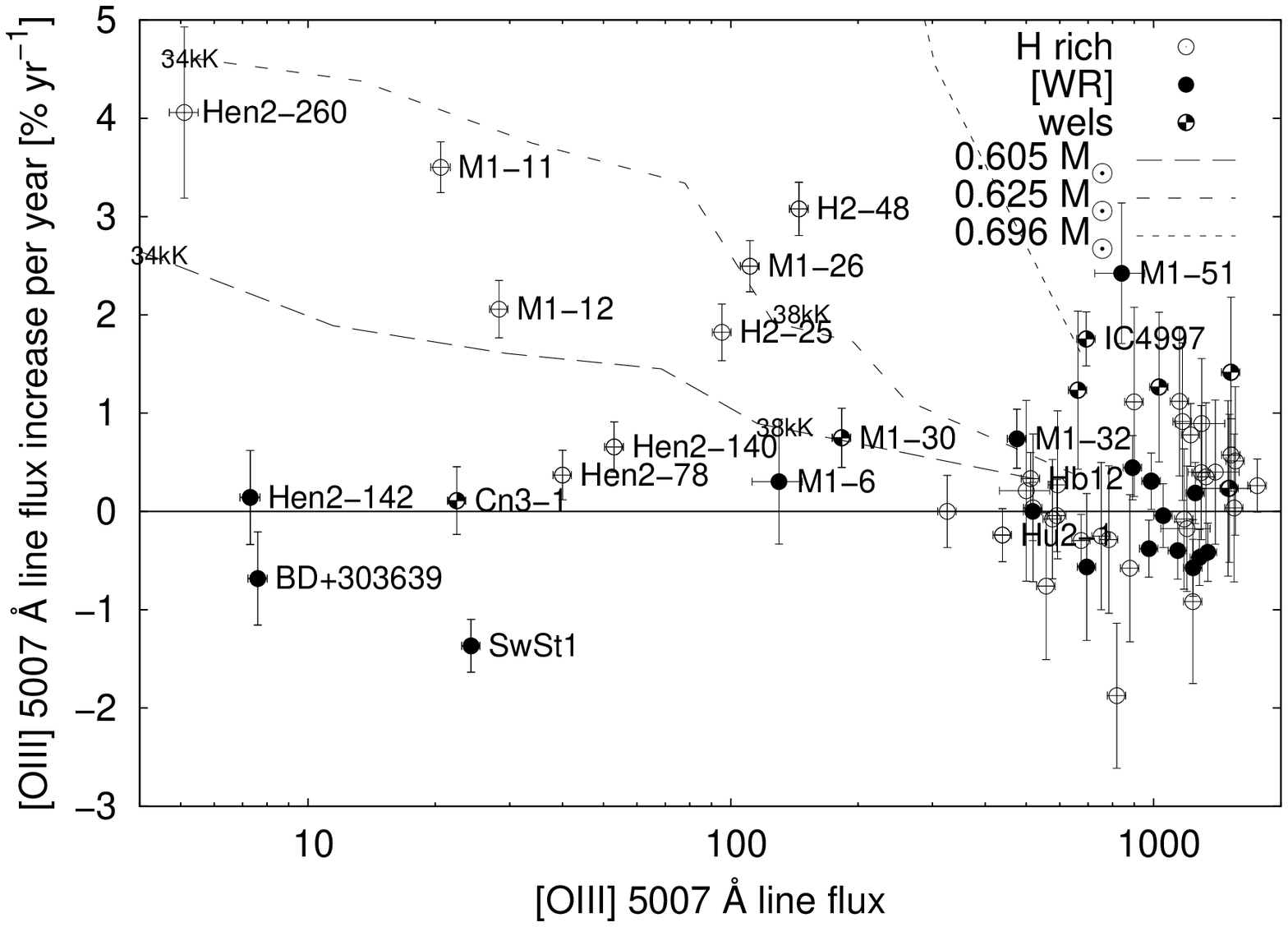}
   \includegraphics[width=8cm]{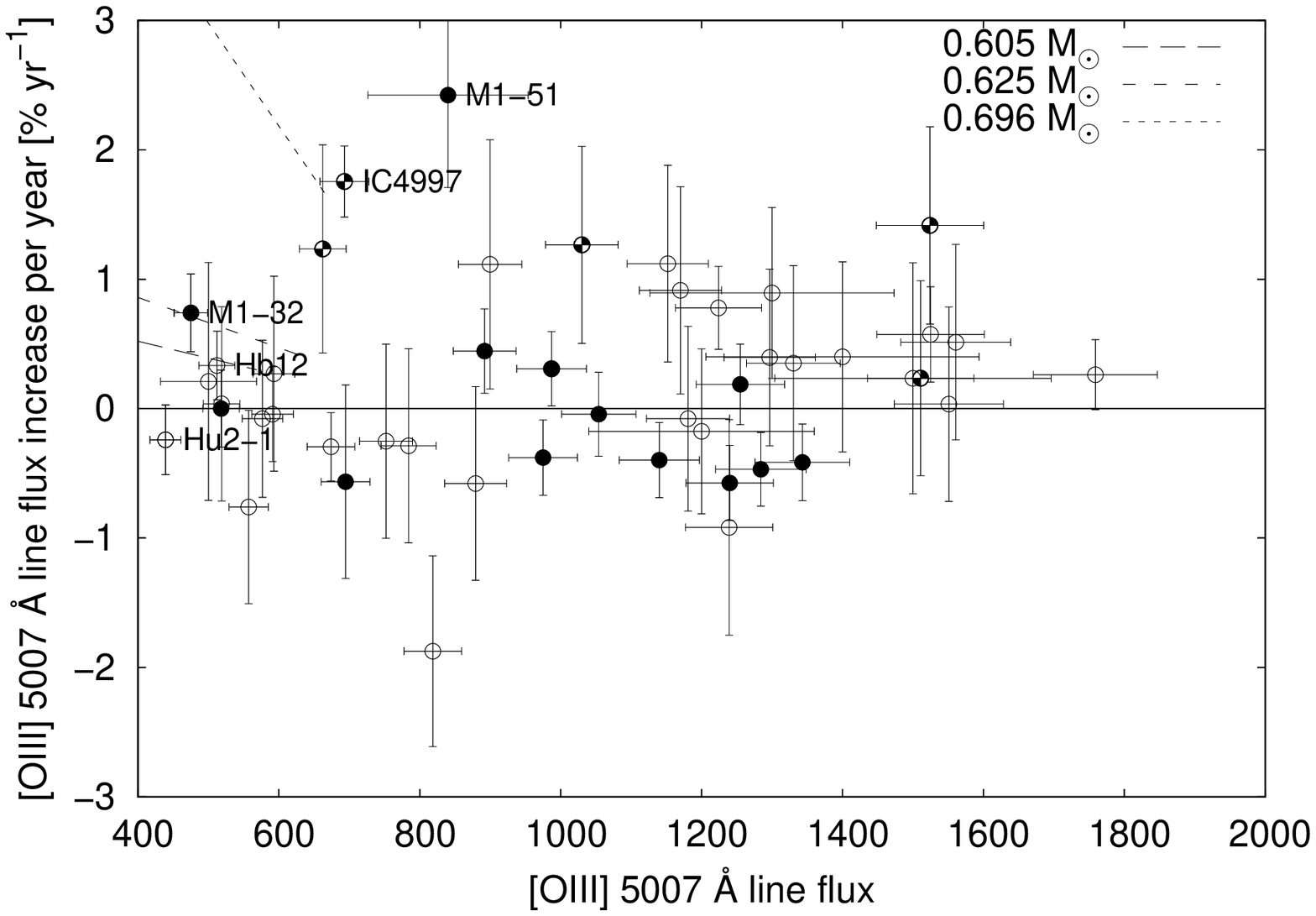}
      \caption{Temporal evolution of the [O\,{\sc iii}] 5007\,\AA\ flux in PNe (normalized to $F \rm (H\beta) = 100$). Cloudy models showing the flux evolution of a PN with a $\rm 0.605$, $\rm 0.625$, and $\rm 0.696 \, M_{\odot}$ (H-burning) core are shown \citep{1995A&A...299..755B}. PNe containing H-rich, $wels$, and [WR] stars are marked with different symbols.
              }
         \label{o3}
   \end{figure}

  \begin{figure}
   \centering
   \includegraphics[width=8cm]{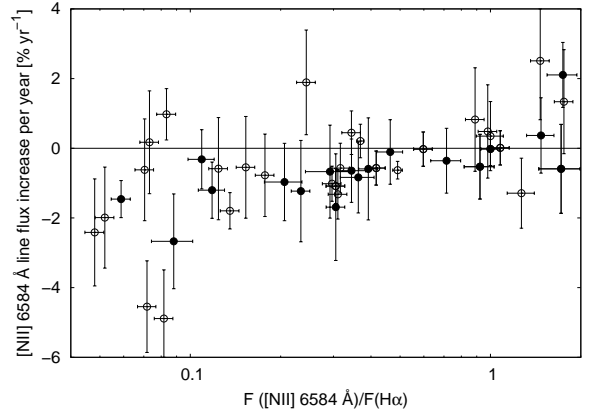}
      \caption{Temporal evolution of the [N\,{\sc ii}] 6584\AA\ flux in PNe (normalized to $F \rm (H\alpha) = 1$). PNe containing $wels$ and [WR] stars are marked with filled circles.
              }
         \label{n2}
   \end{figure}

\begin{table*}
\caption{PNe observed at the SAAO and Toru\'{n} observatories. \label{ratios}}
\begin{center}
{\tiny
\begin{tabular}{ccccccccc}
\hline \hline
name	&date	&wav. range [\AA]&resolution [\AA]&telescope&exp time [s]	&slit ['']&airmass&standard stars\\ \hline
Sp 4-1&2009-05-13&3200-7300&10&Toru\'{n}&1800&4&1.16&Vega, $\rm \iota \, Cyg$\\
M 1-73&2009-05-14&3200-7300&10&Toru\'{n}&1200&4&1.65&$\rm \iota \, Cyg$\\
Cn 3-1	&2009-05-17&3200-7300&10&Toru\'{n}	&600&4&1.20&HR\,7040\\
Hen 2-260&2012-05-07&4300-7400&4&SALT	&900	&1&1.24&G\,93-48\\
H 2-25	&2012-08-21&4300-7400&4&SALT	&1200	&1&1.28&G\,93-48\\
H 2-25	&2013-07-21&3500-7500&6&Radcliffe	&$3 \times 1200$&2&1.07&EG\,21\\
Hen 2-140&2013-07-21&3500-7500&6&Radcliffe&$2 \times 1200$&2&1.22&EG\,21\\
Hen 2-142&2013-07-21&3500-7500&6&Radcliffe&$2 \times 1200$&2&1.35&EG\,21\\
SwSt 1	&2013-07-21&3500-7500&6&Radcliffe	&$2 \times 1200$&2&1.18&EG\,21\\
M 1-26	&2013-07-21&3500-7500&6&Radcliffe	&1200		&2&1.53&EG\,21\\
Hen 2-78&2013-07-22&3500-7500&6&Radcliffe	&$2 \times 1200$&2&1.48&EG\,274\\
H 2-48	&2013-07-22&3500-7500&6&Radcliffe	&120		&2&1.04&EG\,274\\
Hu 2-1	&2013-09-06&3400-7700&13&Toru\'{n}	&$4 \times 600$&4&1.25&Vega\\
BD$+30 3639$&2013-09-06&3400-7700&13&Toru\'{n}	&$2 \times 600$&4&1.24&Vega\\
NGC 6833&2013-09-06&3400-7700&13&Toru\'{n}	&$4 \times 600$&4&1.18&Vega\\
IC 5117	&2013-09-06&3400-7700&13&Toru\'{n}	&$3 \times 600$&4&1.11&Vega\\
Hb 12	&2013-09-06&3400-7700&13&Toru\'{n}	&$3 \times 600$&4&1.04&Vega\\
Cn 3-1	&2013-09-07&3400-7700&13&Toru\'{n}	&$2 \times 600$&4&1.37&Vega\\
IC 4997	&2013-09-07&3400-7700&13&Toru\'{n}	&$2 \times 600$&4&1.27&Vega\\
M 1-11	&2013-11-09&4300-7400&7&SALT	&$2 \times 600$&2&1.22&HILT\,600\\
M 1-12	&2013-12-26&4300-7400&7&SALT	&$2 \times 600$&2&1.20&EG\,21\\
\hline
\end{tabular}}
\end{center}
\end{table*}

\begin{table*}
\caption{Evolution of the [O\,{\sc iii}] 5007\,\AA\ line flux in the low-excitation PNe labeled in Figure \ref{o3}. The spectral types are given by \citet{2011A&A...526A...6W}. \label{ratios2}}
\begin{center}
{\tiny
\begin{tabular}{cccccccccccc}
\hline \hline
name	&diam [arcsec]&	$\rm date$&flux	&slit ['']&$\rm date$&flux&slit ['']&$\rm date$&flux&slit ['']&notes\\
\hline
\multicolumn{12}{c}{H-rich central stars} \\
H 2-25	&4.4[1]3.55[3]	&2013-07-21&84.8&2	&2012-08-21&95.3&1.5	&2001-05-20&75.5&	2&sys?\\
	&		&1984-04-30&58	&4x4	&	&	&	&	&	&		&	\\
H 2-48	&2[1]		&2013-07-22&145	&2	&2002-06-10&113	&2	&1985-07-31&63	&	4x4\\
Hb 12	&1[1]10.12[3]	&2013-09-06&512	&4	&1996/1999	&515	&5	
&1986-07-29&468.1&	4x4&B[e]? WN7?\\
Hen 2-140&2.6[1],4.1[2]&2013-07-21&53	&2	&1984-04-27&44	&	4x4\\
Hen 2-78&3[1]		&2013-07-22&40	&2	&1984-04-29&36	&	4x4\\
Hen 2-260&$< 10$[1]1.93[3]&2012-05-29&7.9&1	&2012-05-07	&7.8	&1.5	&2001-07-12&5.1	&	2&O\\
Hu 2-1	&2.6[1]10.34[3]	&2013-09-06&439	&4	&2001-08&428.8	&	1&	1986-08-01&468	&	4x4\\
M 1-11	&0[1]5.2x5.1[2]2.07[3]&2013-11-09&20.6&2
&2007-02&18&2&1992/1993&11.1&1.5x3.4&emission line\\
	&		&1986-01-18&10	&4x4	&1986-01-09&7.9	&8	&	&	&		&	\\
M 1-12	&0[1]4.35[3]	&2013-12-26&28.3	&2	&2007-02&25&2&1992/1993&20.6&1.5x3.4&emission line\\
	&		&1986-01-18&15	&4x4	&1978-10-10				&13	&1	&	&	&		&	\\
M 1-26	&4.2[1],6.4x6[2]5.13[3]&2013-07-21&111	&2	&1984-04-30&55	&	4x4&		&	&	&B[e]? Of(H)\\
NGC 6833&2[1]0.82[3]	&2013-09-06&674	&4	&1986-07-29&730	&	4x4&	&	&	&Of\\
\\
\multicolumn{12}{c}{$wels$} \\
Cn 3-1	&4.5[1]5.7x4.6[2]&2009-05-17&22.5&4	&2001-08	&21.41	&1	&1988-06-16&22	&4x7.7&wels\\
IC 4997	&1.6[1]7.2[3]	&2013-09-07&703	&4	&1986-07-12&437.0&4x4	&	&	&	&wels\\
M 1-30	&$< 5$[1]	&2010-06-06&183	&1	&1986-07-12&153	&4x4	&	&	&	&wels\\
\\
\multicolumn{12}{c}{WR central stars} \\
BD$+30 3639$&7.5[1]	&2013-09-06&7.61&4	&1988-06-18&9	&4x7.7	&	& 	&	&[WC9]\\
Hen 2-142&3.6[1],4.2x3.1[2]&2013-07-21&7.3&2	&1983-05-03&7	&4x4	&1978-04-26	&15	&1.8	&[WC9]\\
M 1-32	&7.6[1]9.1x8.0[2]&2010-06-06&526&1	&1986-07-29&398	&4x4	&	&	&	&[WC4]pec\\
M 1-51	&9.5[1]		&2004-08&840	&5	&1985-07-27&533	&4x4	&	&	&	&[WO4]pec\\
M 1-6	&$< 5$[1],5.76[3]&2007-02	&130&2	&1986-01-18&122	&4x4	&	&	&	&[WC10-11]?\\
SwSt 1	&$< 5$[1],5.6x5.2[2]&2013-07-21&24.3&2	&1986-07-13&35	&4x4	&	&	&	&[WC9]pec\\
\hline
\end{tabular}
}
\end{center}
{\small
References:
$[1]$ \citet{1992secg.book.....A}, $[2]$ \citet{2003A&A...405..627T} $[3]$ \citet{2011AJ....141..134S}
}
\end{table*}

We analyzed the evolution of the [O\,{\sc iii}] 5007\,\AA /$\rm H \beta$ and
[N\,{\sc ii}] 6584\,\AA/$\rm H\alpha$ line flux ratios in PNe. The observed
[O\,{\sc iii}] 5007\,\AA /$\rm H \beta$ line flux ratio is sensitive to the
temperature of the central star and can be used to trace its evolution. As the
stellar temperature increases, the excitation of the nebula and the [O\,{\sc
iii}] 5007\,\AA /$\rm H \beta$ line flux ratio increase.

The [N\,{\sc ii}] 6584\,\AA/$\rm H\alpha$ flux ratio is less sensitive to the
central star temperature evolution than the [O\,{\sc iii}] 5007\,\AA/$\rm
H\beta$ ratio since the ionization potentials of $\rm N^0$ and H are nearly
equal (13.6\,eV and 14.5\,eV, compared with 35.1\,eV for $\rm O^+$).

We determined the change of the [O\,{\sc iii}] 5007\,\AA/$\rm H\beta$ line flux
ratio between observations taken at two different times in percent per year in
Figure~\ref{o3} and the change of the [N\,{\sc ii}] 6584\,\AA/$\rm H\alpha$ flux
ratio in Figure~\ref{n2}.

We did not correct the observed line fluxes for interstellar extinction. The
extinction for the PNe listed in Table \ref{ratios} range from $c(\rm H\beta) =
0$ for Sp\,4-1 to 2 for H\,2-25 and Hen\,2-140, with an average value of about 1
for our sample. The extinction correction would not affect the flux increase and
only slightly affects the flux ratios (e.g., by 20\% for $c(\rm H\beta) = 2$)
because the pairs of lines are close to each other in wavelengths.

We used flux uncertainties given by the authors. The error bars in Figure 
\ref{o3} and \ref{n2} correspond to the $\rm 1 \, \sigma$ uncertainties. The
flux uncertainties were estimated using different methods by different authors.

To verify the reliability of the error determinations, we inspected the mutual
agreement of the flux uncertainties between \citet{1992secg.book.....A} and four
other groups. For this purpose, we selected a sample of compact and
high-excitation PNe in common for \citet{1992secg.book.....A} and other authors.
This minimizes the chance of the systematic errors due to measuring different
regions of a PN or the evolutionary change of the nebular fluxes.

We compared the line fluxes between \citet{1992secg.book.....A} and these
authors. The median $\chi^2$ values are given in the Table~\ref{obs}. $\chi^2$
values close to 1 confirm the reliability of the error determinations. However,
deviations stronger than expected occur for some PNe for lines at the blue and
red end of the wavelength range.

We adopted a 5\% $\rm 1 \, \sigma$ uncertainty (or the last significant digit of
the flux value, if larger) for the fluxes published without errors. This is a
typical calibration uncertainty for spectroscopic observations and may be
applied for strong lines (in particular [O\,{\sc iii}] 5007\,\AA, [N\,{\sc ii}]
6584\,\AA\ and hydrogen lines used by us), for which other sources of errors are
less important.

We assumed an error of 5\% for the fluxes of strong lines in the A quality
spectra published by \citet{1996A&A...307..215C} instead of 1\% adopted by the
authors. The 1\% uncertainty did not take the calibration uncertainties into
account

A comparison of the line flux ratios obtained at different times requires
caution. For spatially resolved PNe, the flux line ratios may depend on the
diameter and the surface brightness distribution of a PN, slit width, seeing,
pointing, and the details of the spectrum subtraction procedure.

The observed [N\,{\sc ii}] 6584\,\AA/$\rm H\alpha$ flux ratio may depend on the
projected aperture, because [N\,{\sc ii}] 6584\,\AA\ emission might be
suppressed in the dense regions and in the inner regions of PNe with hot central
stars (because of the ionization of $\rm N^+$). More narrow slits might result
in a lower observed [N\,{\sc ii}] 6584\,\AA/$\rm H\alpha$ flux ratio. However,
this effect seems to be unimportant in our sample, because we do not observe a
significant flux change in any of our PNe within the uncertainties (Figure
\ref{n2}).

The observed [O\,{\sc iii}] 5007\,\AA /$\rm H \beta$ line flux ratio can be
higher when using more narrow slits. More narrow slits or apertures probe the
inner (high-excitation) regions of spatially resolved PNe, which results in
higher observed [O\,{\sc iii}] 5007\,\AA/$\rm H\beta$ flux ratios. This effect
may be particularly important for young PNe that show strong ion
stratification, for which the radii of the $\rm O^{++}$ zones may be
significantly smaller than the outer radii of the $\rm H^+$ zones.

More recent surveys tend to use smaller slits or apertures than older surveys.
Almost all of the observed PNe have less than 10 arcsec in diameter, but many of
them were resolved by the ground-based telescopes. Narrower slits can result in
deriving higher [O\,{\sc iii}] 5007\,\AA /$\rm H \beta$ line flux ratios and can
mimic an excitation increase of a PN. 

We studied the influence of a finite slit width in PNe that showed a change of
the observed [O\,{\sc iii}] 5007\,\AA/$\rm H\beta$ flux ratio. We used available
Hubble Space Telescope (HST) images \citep{2011AJ....141..134S} for this
purpose. Most of the HST observations used the $\rm H\alpha$ filter. Since we do
not know the [O\,{\sc iii}] 5007\,\AA\ line surface brightness distribution for
our PNe (either because HST observations in the [O\,{\sc iii}] 5007\AA\ line
filter were not performed, or the S/N ratio was very poor), we assumed all the
[O\,{\sc iii}] 5007\AA\ line emission to originate from an unresolved region in
the center of a PN. In this case, the finite slit width would affect the
[O\,{\sc iii}] 5007\,\AA/$\rm H\beta$ flux ratio most.

We convolved the $\rm H\alpha$ HST images with a Gaussian with the full width at
half maximum (FWHM) equal to 2 arcsec to account for the seeing. The [O\,{\sc
iii}] 5007\AA\ line brightness distribution convolved with the seeing produced a
2D Gaussian with the FWHM of 2 arcsec. Then we integrated the [O\,{\sc iii}]
5007\,\AA\ and $\rm H\alpha$ line flux along a slit of different diameters
passing the nebula in the N-S orientation. The fluxes were normalized to unity
so that the [O\,{\sc iii}] 5007\,\AA/$\rm H\beta$ flux ratio would be equal to 1
for an infinite aperture. The observed [O\,{\sc iii}] 5007\,\AA/$\rm H\beta$
flux ratio as a function of the slit width is shown in Figure \ref{m111} for
different PNe.

For M\,1-11 and M\,1-12 a narrow slit would increase the [O\,{\sc iii}]
5007\,\AA/$\rm H\beta$ flux ratio by 15-20\% compared with an infinite aperture
under 2 arcsec seeing. The same ratio would be increased by 40\% in M\,1-26 and
by 60\% in H\,2-25. Depending on the time span of the observations made with
different slits, using different slits could easily mimic a significant increase
of the excitation of the PN.

The resolution of the spectra can also affect the derived line fluxes. The
[O\,{\sc iii}] 5007\,\AA\ line may be blended with the [Fe\,{\sc iii}]
5011\,\AA\ {and He\,{\sc i} 5015\,\AA\ lines} in observations with resolutions
of about 10\,\AA\ or lower.

Finally, the observed flux ratios may be affected by the differential
atmospheric refraction, as some of the observations were made with the slit
oriented in the east-west direction. However, we find this influence negligible
for the [O\,{\sc iii}] 5007\,\AA\ to $\rm H\beta$ line ratio: at an airmass of
2, the relative displacement would only be 0.12 arcsec for the respective
wavelengths.

\subsection{PNe with H-rich central stars}

Only a few PNe have $F {\rm ([O\,III]) 5007\,\AA} \leq 2 \times F {\rm
(H\beta)}$, corresponding to a stellar temperature of up to 40\,kK. However, the
evolution of the nebular flux is evident for most of them. We determined whether
this might be the result of the different slit widths or apertures used
individually for each object as each object has different diameter, surface
distribution, and was observed with a different set of apertures.

PNe H\,2-48 and Hen\,2-260 are unresolved in ground-based observations
(Table~\ref{ratios2}). An increase of the [O\,{\sc iii}] 5007\,\AA\ flux is
clearly observed in both cases and most likely reflects an increase of the
temperature of the central stars.

\begin{figure}
   \centering
   \includegraphics[width=4.25cm]{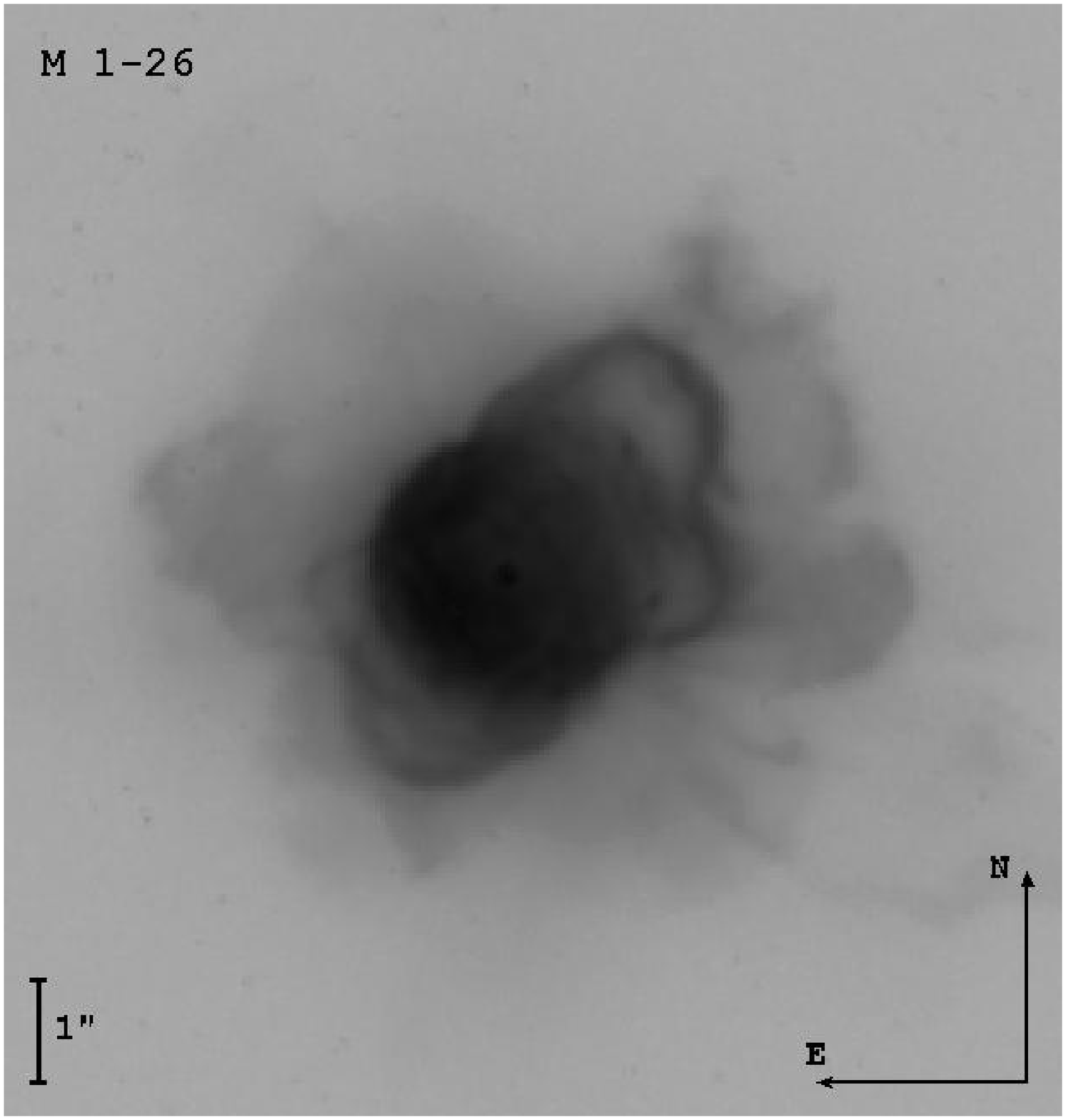}
   \includegraphics[width=4.25cm]{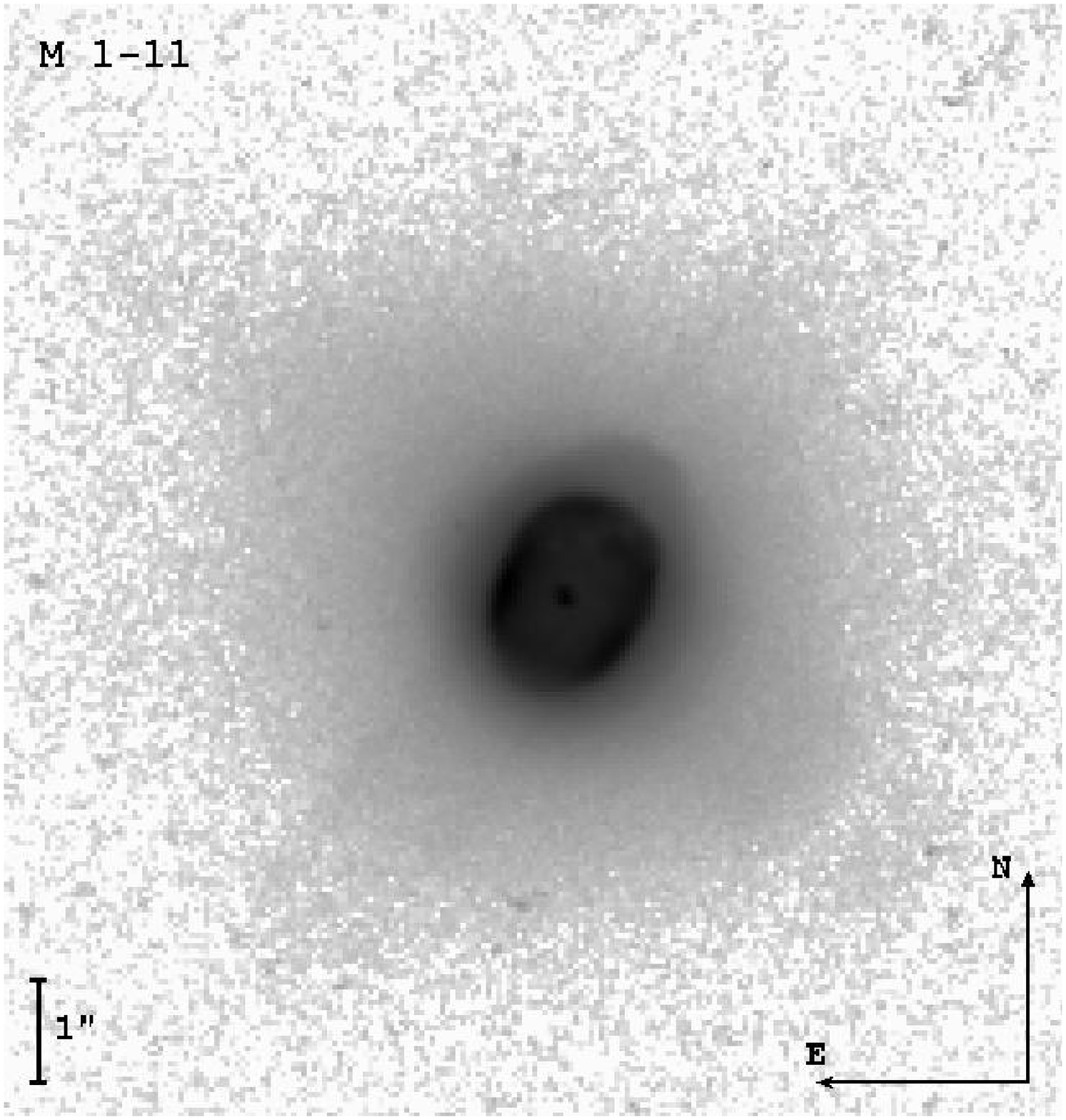}
   
   \includegraphics[width=4.25cm]{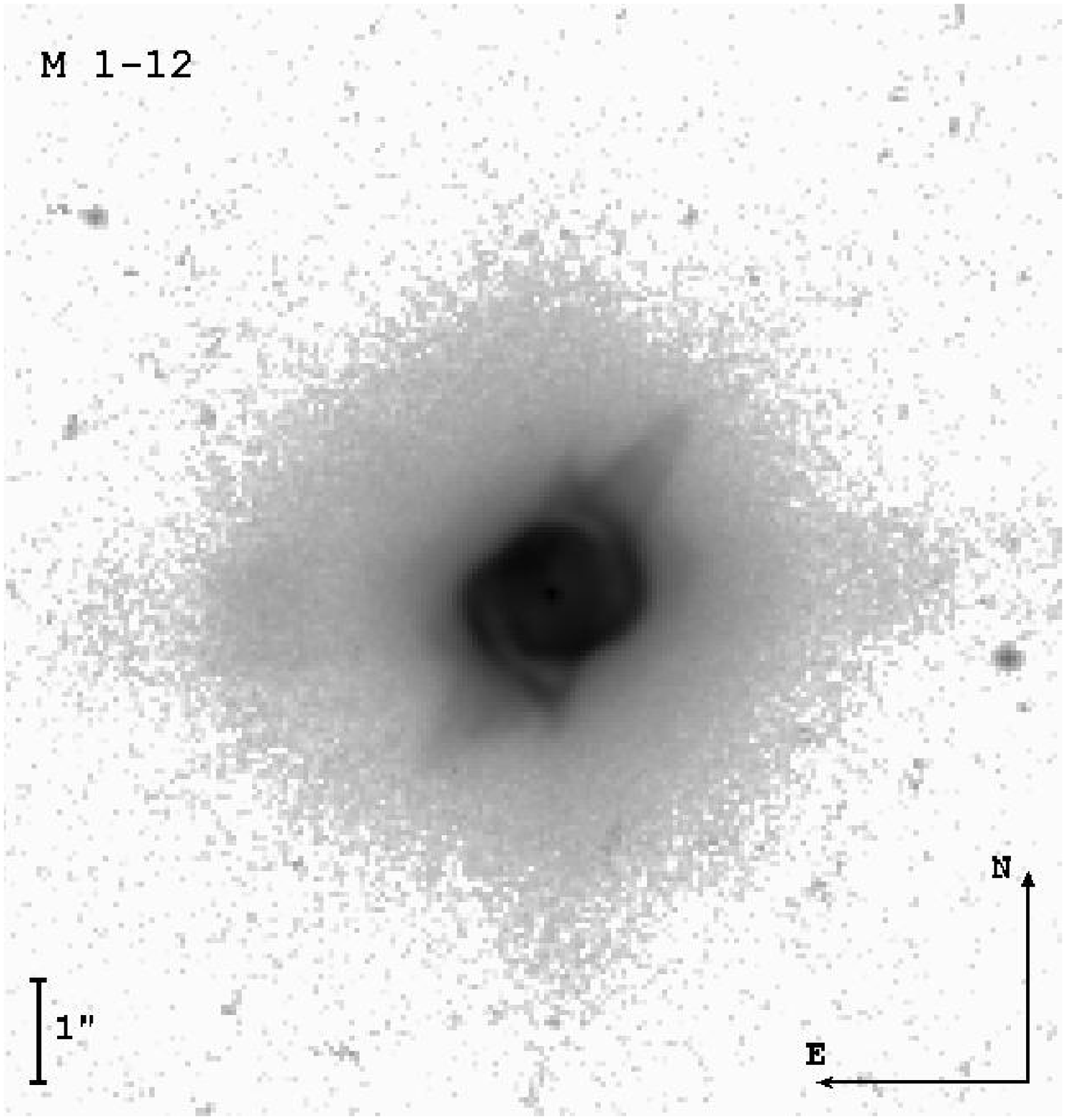}
   \includegraphics[width=4.25cm]{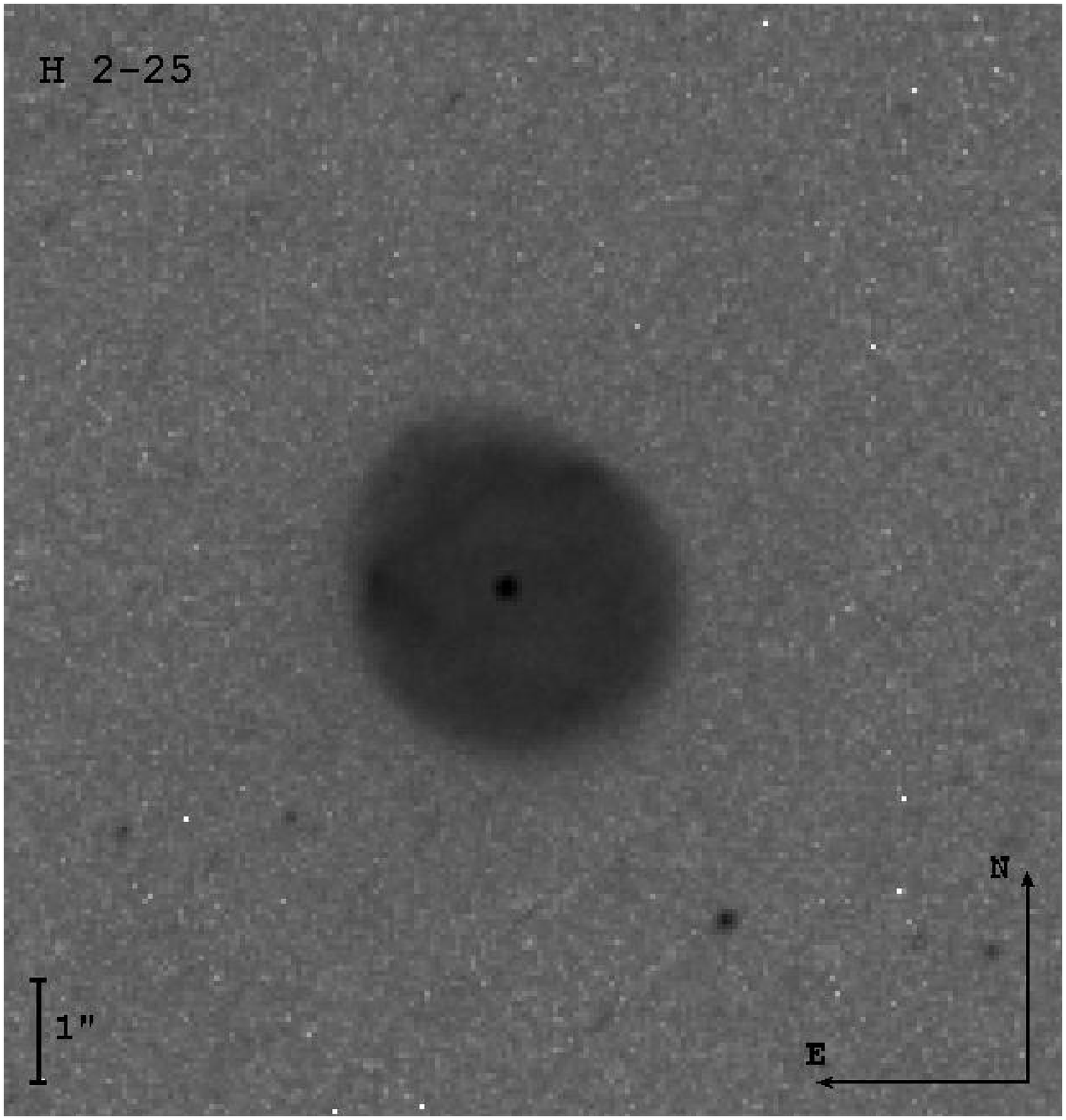}
      \caption{$\rm H\alpha$ HST image of the PN M\,1-26. Log stretch and reverse gray scale are applied.
              }
         \label{m126hst}
   \end{figure}

In M\,1-26 the observed [O\,{\sc iii}] 5007\,\AA\ to the $\rm H\beta$ line flux
ratio has doubled between 1984 and 2013. The first observation used a 4 arcsec
aperture, the second a 2 arcsec slit, while the PN has a diameter of about 10
arcsec in $\rm H \alpha$. However, the bright central region of a diameter of
about 2.5 arcsec contributes most of the nebular flux to the $\rm H \alpha$ line
(Figure \ref{m126hst}). A group of faint multipolar lobes and an extended halo
have low surface brightness and contribute negligibly to both observations.

\begin{figure}
   \centering
   \includegraphics[width=8cm]{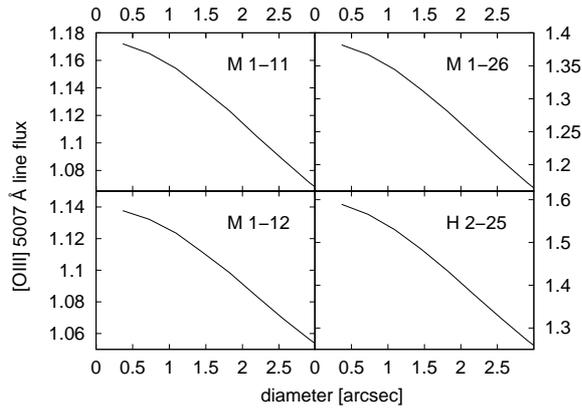}
      \caption{Dependence of the observed [O\,{\sc iii}] 5007\AA/$\rm H\beta$ line flux ratio on the slit width.
              }
         \label{m111}
   \end{figure}

Using different slits might increase the observed [O\,{\sc iii}] 5007\,\AA/$\rm
H\beta$ ratio by up to 18\% between 1984 and 2013 (corresponding to the rate of
$\rm 0.6\% \, yr^{-1}$) under 2 arcsec seeing (Figure \ref{m111}). The seeing
during the 2013 observation was even worse than assumed (about 3 arcsec).
Different slit widths cannot account for the the observed increase of the
[O\,{\sc iii}] 5007\,\AA/$\rm H\beta$ ratio between 1984 and 2013.

We extracted our spectrum taken in 2013 using different apertures in the spatial
axis. We obtained the [O\,{\sc iii}] 5007\AA/$\rm H\beta$ ratio (normalized to 
$F {\rm (H\beta)} = 100$) of 111 using the 10 arcsec aperture and 130 using the
aperture of 2 arcsec. This confirms that the [O\,{\sc iii}] 5007\AA\ emission
originates from a smaller region than the $\rm H\beta$ emission.

\begin{figure}
   \centering
   \includegraphics[width=8cm]{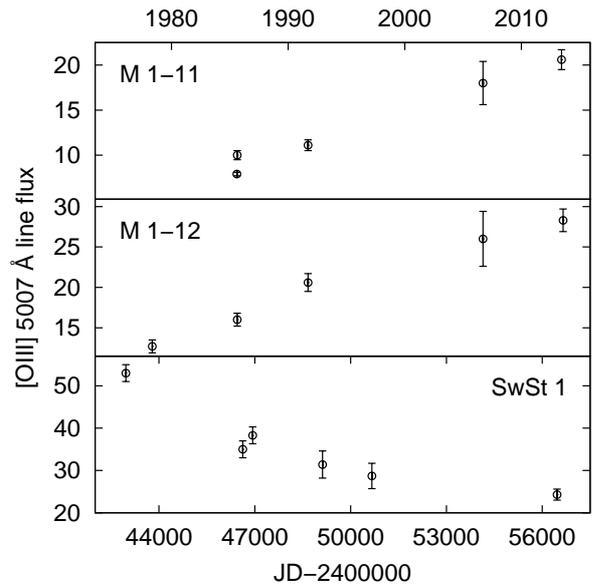}
      \caption{[O\,{\sc iii}] 5007\AA\ flux evolution in the PN M\,1-11, M\,1-12 and SwSt\,1.
              }
         \label{m111flx}
   \end{figure}

M\,1-11 and M\,1-12 were observed five times and show a gradually increasing
[O\,{\sc iii}] 5007\,\AA\ line flux in time (Figure \ref{m111flx}).

The flux evolution in M\,1-11 was previously reported by
\citet{2005A&AT...24..291K}, although she did not report the flux evolution in
M\,1-12.

The compact PN M\,1-11 emits most of the $\rm H\alpha$ flux from the central 2
arcsec (Figure \ref{m126hst}). For a 2 arcsec seeing, a 2 arcsec slit would
increase the observed [O\,{\sc iii}] 5007\AA/$\rm H\beta$ ratio by only up to
10\% (Figure \ref{m111}), corresponding to an increase rate of the flux ratio
lower than $\rm 0.3\% \, yr^{-1}$. This cannot account for the observed flux
increase between 1986 and 2013.

We compared the [O\,{\sc iii}] 5007\AA/$\rm H\beta$ ratio obtained from the 2013
SALT spectrum extracted using different apertures in the spatial axis, but we
obtained similar results. This indicates that the PN was not spatially resolved
in our observation; alternatively, the [O\,{\sc iii}] 5007\AA\ and $\rm H\beta$
line brightness distributions are similar. The seeing was about 2.5 arcsec
during the observation.

The HST image of M\,1-12 shows a bright inner region of about 2 arcsec in
diameter and a faint halo extending up to 4 arcsec in diameter, but contributing
less than 10\% of the total $\rm H \alpha$ flux of the PN (Figure
\ref{m126hst}). 

A 2 arcsec slit for 2 arcsec seeing would increase the observed [O\,{\sc iii}]
5007\,\AA/$\rm H\beta$ line flux ratio by only up to 10\%, corresponding to an
increase rate lower than $\rm 0.3\% \, yr^{-1}$. This cannot account for the
changes observed between 1986 and 2013.

The oldest observation was made in 1978 by \citet{1994MNRAS.271..257K}. It used
the most narrow slit of all the spectra obtained for M\,1-12 and shows the
lowest [O\,{\sc iii}] 5007\,\AA/$\rm H\beta$ ratio. This is opposite to what
would be expected if there had been no change of the [O\,{\sc iii}]
5007\,\AA/$\rm H\beta$ ratio in the PN since 1978.

The [O\,{\sc iii}] 5007\AA/$\rm H\beta$ line flux ratio was not sensitive to the
width of the aperture used for the extraction of the SALT spectrum taken in
2013.

H\,2-25 showed an increase of the [O\,{\sc iii}] 5007\,\AA /$\rm H\beta$ line
flux ratio by almost 100\% between 1984 and 2013. The nebula has almost
spherical appearance with a uniform brightness distribution in the HST $\rm H
\alpha$ image (Figure \ref{m126hst}). The [O\,{\sc iii}] 5007\,\AA\ line image
is also available, but the extended emission was too faint to be detected.

The analysis of the HST image implies that a 2 arcsec slit would increase the
observed [O\,{\sc iii}] 5007\,\AA /$\rm H \beta$ flux ratio by up to 26\%
(corresponding to an increase rate of $\rm 0.8\% \, yr^{-1}$) compared with the
average flux for 2 arcsec seeing. This accounts for a quarter of the change
observed between 1984 and 2013. Most of the [O\,{\sc iii}] 5007\,\AA\ line flux
increase appears to be physical.

The SALT spectrum confirms, that the [O\,{\sc iii}] 5007\,\AA\ emission is more
compact than the $\rm H \beta$ emission in H\,2-25. We obtained a 105.0 [O\,{\sc
iii}] 5007\,\AA\ line flux integrated over 10 arcsec aperture compared with 84.8
using a narrower aperture of 0.5 arcsec in the spatial axis. The FWHM of the
stellar profile was 1.7 arcsec, 2.9 arcsec for the hydrogen lines and 2.2 for
the [O\,{\sc iii}] and He\,{\sc i} lines.

Hen\,2-140 and Hen\,2-78 are large enough to be spatially resolved by
ground-based instruments. The weak observed [O\,{\sc iii}] 5007\,\AA\ line flux
increase can easily be explained by the different apertures used, since more
recent observations used smaller slit widths which probed the inner regions of
the PNe.

\subsection{PNe with weak emission line central stars}

IC\,4997 shows a set of extended, low surface brightness multipolar lobes that
surround a bright, compact center that is saturated in the HST image. The
observed increase in the [O\,{\sc iii}] 5007\,\AA /$\rm H\beta$ line flux ratio
is most likely physical although different slits or aperture sizes were used.
The spectrum evolution was also confirmed by \citet{2009ARep...53.1155K}, who
monitored the object for forty years. The heating rate appears to be relatively
high in this star.

Cn\,3-1 was observed using different slit widths and apertures and always showed
a similar flux, thus there is no indication for a flux change in this object.

M\,1-30 shows an increase of the [O\,{\sc iii}] 5007\,\AA /$\rm H\beta$ line
flux ratio in time. However, it was observed with a 4x4 arcsec aperture and then
using a 1 arcsec slit. The PN might be spatially resolved in the latter
observation, thus the observed flux increase requires further confirmation.

\subsection{WR-type central stars}

None of the PNe containing late-type Wolf-Rayet (WR) stars showed an increase of
the [O\,{\sc iii}] 5007\,\AA /$\rm H\beta$ line flux ratio. Instead,
SwSt\,1 showed a decreasing [O\,{\sc iii}]\,5007\,\AA\ flux between our
observation and the spectrum taken in 1986 by \citet{1992secg.book.....A}.

There are more spectra of this well-studied object published in the literature.
The temporal evolution of the [O\,{\sc iii}]\,5007\,\AA\ line flux is shown in
Figure \ref{m111flx}. \citet{1984MNRAS.206..293F} reported the reddened [O\,{\sc
iii}]\,5007\,\AA\ line flux of $\rm 53$ for their observation in 1976, the
highest flux ever measured for this line in SwSt\,1. We obtained an [O\,{\sc
iii}]\,5007\,\AA\ line flux of $24.3$, which is the lowest value reported so
far.

We marginally resolved the He\,{\sc i} 5015\,\AA\ line. The [O\,{\sc iii}] 5007
\AA\ / 4959 \AA\ line flux ratio of 3.11 close to the theoretical ratio of 2.98
confirms that the deblending of the [O\,{\sc iii}] 5007 \AA\ and He\,{\sc i}
5015\,\AA\ lines is reliable. The He\,{\sc i} 5015\,\AA\ line only contributes
about 10\% to the [O\,{\sc iii}] 5007\,\AA\ line flux. The observed decrease of
the [O\,{\sc iii}] 5007 \AA\ is real, although it may be slightly affected by
blending with He\,{\sc i} 5015\,\AA\ in some of the spectra (e.g.,
\citet{1992secg.book.....A}).

\citet{2001MNRAS.328..527D} studied historical spectroscopy of this object and
did not report significant changes. However, their analysis was based on the
qualitative analysis of the nebular iron and stellar emission lines.

We did not detect a significant decrease of the [O\,{\sc iii}] 5007\,\AA\ line
flux in $\rm BD+30 \, 3639$.

Another PN containing a WR-type central star is M\,1-32. It shows an increase of
the [O\,{\sc iii}] 5007\,\AA\ /$\rm H\beta$ line flux ratio in time. However,
the aperture used in the latter observation was only 1 arcsec, and the PN might
be spatially resolved, so this case requires confirmation.

M\,1-51 is a PN containing an intermediate-type [WR] star. Both observations
used similar apertures, so the observed flux increase is most likely physical.
M\,1-51 has a relatively strong [O\,{\sc iii}] 5007\,\AA\ line, and the
observations may suggest a relatively fast heating rate for this object.

\subsection{Modeling}

We modeled the evolution of the [O\,{\sc iii}] 5007\,\AA /$\rm H\beta$ line flux
ratio in PNe with time using the post-AGB evolutionary tracks and exploratory
Cloudy c10.01 models \citep{1998PASP..110..761F}. We chosen the evolutionary
tracks for the final masses of 0.605, 0.625 and 0.696\,$\rm M_{\odot}$ computed
by \citet{1995A&A...299..755B}. We ran a set of photoionization models for each
track. Luminosity and temperature changed with age according to the evolutionary
models. The temperatures and luminosities of the central star were interpolated
(in the log scale) to achieve better coverage in time.

The photoionization models assumed spherical symmetry. The inner radius was
assumed to expand at a constant velocity of $\rm 25\,km\,s^{-1}$. The nebular
density (constant across the nebular shell) decreased with the cube of
time. Inner radius and nebular density were set to $\rm 16.8\,cm$ and $\rm
4.4\,cm^{-3}$ in log for the $\rm T_{eff}$ of 30,000\,K. We used an electron
density of $\rm 1 \, cm^{-3}$ as a stopping criterion. The nebular abundances
were taken from \citet{1983ApJS...51..211A} and \citet{1989SSRv...51..339K}.

We computed the increase of the [O\,{\sc iii}] 5007\,\AA\ line flux per year for
the model pairs. The observed evolution of the [O\,{\sc iii}] 5007\,\AA\ line
flux in PNe is plotted along with the model computed for different final masses
in Figure~\ref{o3}. The stellar temperatures are marked along the tracks.

TLUSTY model atmospheres were used for the stellar radiation of the H-rich
central stars \citep{1995ApJ...439..875H}. The evolution of the He-burning
models along the horizontal part of the Herzsprung-Russell diagram is
approximately three times slower than the H-burning models computed by
\citet{1995A&A...299..755B}.

\section{Summary}

We studied the evolution of the central stars of PNe through the variability of
nebular fluxes. All PNe hosting H-rich central stars with a flux ratio $F {\rm
([O\,III]) 5007\,\AA} \leq 2 \times F {\rm (H\beta)}$ show an increase of the
[O\,{\sc iii}] 5007\,\AA\ /$\rm H\beta$ line flux ratio. Only in two cases
(Hen\,2-140 and Hen\,2-76) this requires further verification. In the six
remaining cases (Hen\,2-260, M\,1-11, M\,1-12, M\,1-26 H\,2-48 and H\,2-25) the
flux increase is physical at least in part. The pace of the evolution is
consistent with evolutionary models for the central star masses in the range of
$\rm 0.6-0.65\,M_\odot$. Accurate mass determinations for central stars will
require individual modeling.

None of the PNe hosting late-type [WR] stars shows an increase of the [O\,{\sc
iii}] 5007\,\AA\ line flux. H-deficient late-type [WR] stars may form a distinct
evolutionary channel \citep{2008ASPC..391..209D} and their evolution may be much
slower than the evolution of early-type [WR] stars. Perhaps late-type [WR] stars
originate from central stars with lower masses than the earlier-type [WR] stars
\citep{2009A&A...500.1089G}. An alternative scenario would be binary evolution.
Slow evolution is in accordance with a relative high number of the late-type
[WR] stars compared to intermediate and early-type [WR] stars
\citep{2008ASPC..391..209D}.

Quite unexpectedly, we detected the decrease of the [O\,{\sc iii}] 5007\,\AA\
line flux in SwSt\,1, which hosts a late-type [WR] star. A decrease in intensity
of the C\,{\sc iv} 5470\AA\ line was reported by \citet{2001MNRAS.328..527D}.
This suggests an evolutionary change of the stellar wind properties
\citep{1989A&A...210..236S}.

The recombination time-scale of $\rm O^{+}$ appears to be of about 35 years or
shorter, if there is a constant contribution to the [O\,{\sc iii}] 5007\,\AA\
line flux from other parts of the PN. This implies a lower limit of the electron
density of $\rm 200\,cm^{-3}$ \citep{2004A&A...426..145L}. The weakening stellar
wind indicates an evolution of the star to higher temperatures. The [O\,{\sc
iii}] emission might originate from the regions ionized by the stellar wind,
which now is weakening.

One of the intermediate-type [WR] stars shows a rapid evolution of the [O\,{\sc
iii}] 5007\,\AA\ line flux. The [O\,{\sc iii}] 5007\,\AA\ flux increase observed
in another [WR] star requires further verification.

Weak emission line central stars ($wels$) form a distinct evolutionary path
\citep{2006A&A...451..925G}. This group of stars is not uniform and may contain
both H- and He-burning central stars. The PN IC\,4997, hosting a $wels$ central
star, rapidly evolves in time. Cn\,3-1 does not show any significant changes
although it contains a much cooler central star than IC\,4997. The flux increase
in M\,1-30 requires confirmation.

\begin{acknowledgements}

This work was financially supported by NCN of Poland through grants No.
2011/01/D/ST9/05966 and 719/N-SALT/2010/0. PvH acknowledges support from the
Belgian Science Policy office through the ESA PRODEX program. This paper uses
observations made at the South African Astronomical Observatory (SAAO). Some of
the observations reported in this paper were obtained with the Southern African
Large Telescope (SALT). We thank the referee for the comments on the paper.

\end{acknowledgements}

\bibliographystyle{aa} 
\bibliography{csevol}

\begin{thebibliography}{33}
\expandafter\ifx\csname natexlab\endcsname\relax\def\natexlab#1{#1}\fi

\bibitem[{{Acker} {et~al.}(1992){Acker}, {Marcout}, {Ochsenbein}, {Stenholm},
  {Tylenda}, \& {Schohn}}]{1992secg.book.....A}
{Acker}, A., {Marcout}, J., {Ochsenbein}, F., {et~al.} 1992, {The
  Strasbourg-ESO Catalogue of Galactic Planetary Nebulae. Parts I, II.}

\bibitem[{{Aller} \& {Czyzak}(1983)}]{1983ApJS...51..211A}
{Aller}, L.~H. \& {Czyzak}, S.~J. 1983, \apjs, 51, 211

\bibitem[{{Bl\"{o}cker}(1995)}]{1995A&A...299..755B}
{Bl\"{o}cker}, T. 1995, \aap, 299, 755

\bibitem[{{Cuisinier} {et~al.}(1996){Cuisinier}, {Acker}, \&
  {Koeppen}}]{1996A&A...307..215C}
{Cuisinier}, F., {Acker}, A., \& {Koeppen}, J. 1996, \aap, 307, 215

\bibitem[{{De Marco}(2008)}]{2008ASPC..391..209D}
{De Marco}, O. 2008, in Astronomical Society of the Pacific Conference Series,
  Vol. 391, Hydrogen-Deficient Stars, ed. A.~{Werner} \& T.~{Rauch}, 209

\bibitem[{{De Marco} {et~al.}(2001){De Marco}, {Crowther}, {Barlow}, {Clayton},
  \& {de Koter}}]{2001MNRAS.328..527D}
{De Marco}, O., {Crowther}, P.~A., {Barlow}, M.~J., {Clayton}, G.~C., \& {de
  Koter}, A. 2001, \mnras, 328, 527

\bibitem[{{Escudero} {et~al.}(2004){Escudero}, {Costa}, \&
  {Maciel}}]{2004A&A...414..211E}
{Escudero}, A.~V., {Costa}, R.~D.~D., \& {Maciel}, W.~J. 2004, \aap, 414, 211

\bibitem[{{Exter} {et~al.}(2004){Exter}, {Barlow}, \&
  {Walton}}]{2004MNRAS.349.1291E}
{Exter}, K.~M., {Barlow}, M.~J., \& {Walton}, N.~A. 2004, \mnras, 349, 1291

\bibitem[{{Feibelman} {et~al.}(1992){Feibelman}, {Aller}, \&
  {Hyung}}]{1992PASP..104..339F}
{Feibelman}, W.~A., {Aller}, L.~H., \& {Hyung}, S. 1992, \pasp, 104, 339

\bibitem[{{Ferland} {et~al.}(1998){Ferland}, {Korista}, {Verner}, {Ferguson},
  {Kingdon}, \& {Verner}}]{1998PASP..110..761F}
{Ferland}, G.~J., {Korista}, K.~T., {Verner}, D.~A., {et~al.} 1998, \pasp, 110,
  761

\bibitem[{{Flower} {et~al.}(1984){Flower}, {Goharji}, \&
  {Cohen}}]{1984MNRAS.206..293F}
{Flower}, D.~R., {Goharji}, A., \& {Cohen}, M. 1984, \mnras, 206, 293

\bibitem[{{Garc{\'{\i}}a-Rojas} {et~al.}(2012){Garc{\'{\i}}a-Rojas},
  {Pe{\~n}a}, {Morisset}, {Mesa-Delgado}, \& {Ruiz}}]{2012A&A...538A..54G}
{Garc{\'{\i}}a-Rojas}, J., {Pe{\~n}a}, M., {Morisset}, C., {Mesa-Delgado}, A.,
  \& {Ruiz}, M.~T. 2012, \aap, 538, A54

\bibitem[{{Gesicki} {et~al.}(2006){Gesicki}, {Zijlstra}, {Acker}, {G{\'o}rny},
  {Gozdziewski}, \& {Walsh}}]{2006A&A...451..925G}
{Gesicki}, K., {Zijlstra}, A.~A., {Acker}, A., {et~al.} 2006, \aap, 451, 925

\bibitem[{{G{\'o}rny} {et~al.}(2009){G{\'o}rny}, {Chiappini}, {Stasi{\'n}ska},
  \& {Cuisinier}}]{2009A&A...500.1089G}
{G{\'o}rny}, S.~K., {Chiappini}, C., {Stasi{\'n}ska}, G., \& {Cuisinier}, F.
  2009, \aap, 500, 1089

\bibitem[{{Hajduk} {et~al.}(2014){Hajduk}, {van Hoof}, {Gesicki}, {Zijlstra},
  {G{\'o}rny}, \& {G{\l}adkowski}}]{2014arXiv1405.0800H}
{Hajduk}, M., {van Hoof}, P.~A.~M., {Gesicki}, K., {et~al.} 2014, ArXiv
  e-prints

\bibitem[{{Heap}(1993)}]{1993IAUS..155..484H}
{Heap}, S.~R. 1993, in IAU Symposium, Vol. 155, Planetary Nebulae, ed.
  R.~{Weinberger} \& A.~{Acker}, 484

\bibitem[{{Heap} \& {Altner}(1993)}]{1993AAS...18310104H}
{Heap}, S.~R. \& {Altner}, B. 1993, in Bulletin of the American Astronomical
  Society, Vol.~25, American Astronomical Society Meeting Abstracts, 1441

\bibitem[{{Henry} {et~al.}(2010){Henry}, {Kwitter}, {Jaskot}, {Balick},
  {Morrison}, \& {Milingo}}]{2010ApJ...724..748H}
{Henry}, R.~B.~C., {Kwitter}, K.~B., {Jaskot}, A.~E., {et~al.} 2010, \apj, 724,
  748

\bibitem[{{Hubeny} \& {Lanz}(1995)}]{1995ApJ...439..875H}
{Hubeny}, I. \& {Lanz}, T. 1995, \apj, 439, 875

\bibitem[{{Kaler} {et~al.}(1996){Kaler}, {Kwitter}, {Shaw}, \&
  {Browning}}]{1996PASP..108..980K}
{Kaler}, J.~B., {Kwitter}, K.~B., {Shaw}, R.~A., \& {Browning}, L. 1996, \pasp,
  108, 980

\bibitem[{{Khromov}(1989)}]{1989SSRv...51..339K}
{Khromov}, G.~S. 1989, \ssr, 51, 339

\bibitem[{{Kingsburgh} \& {Barlow}(1994)}]{1994MNRAS.271..257K}
{Kingsburgh}, R.~L. \& {Barlow}, M.~J. 1994, \mnras, 271, 257

\bibitem[{{Kondratyeva}(2005)}]{2005A&AT...24..291K}
{Kondratyeva}, L.~N. 2005, Astronomical and Astrophysical Transactions, 24, 291

\bibitem[{{Kostyakova} \& {Arkhipova}(2009)}]{2009ARep...53.1155K}
{Kostyakova}, E.~B. \& {Arkhipova}, V.~P. 2009, Astronomy Reports, 53, 1155

\bibitem[{{Kwitter} {et~al.}(2003){Kwitter}, {Henry}, \&
  {Milingo}}]{2003PASP..115...80K}
{Kwitter}, K.~B., {Henry}, R.~B.~C., \& {Milingo}, J.~B. 2003, \pasp, 115, 80

\bibitem[{{Lechner} \& {Kimeswenger}(2004)}]{2004A&A...426..145L}
{Lechner}, M.~F.~M. \& {Kimeswenger}, S. 2004, \aap, 426, 145

\bibitem[{{Sahai} {et~al.}(2011){Sahai}, {Morris}, \&
  {Villar}}]{2011AJ....141..134S}
{Sahai}, R., {Morris}, M.~R., \& {Villar}, G.~G. 2011, \aj, 141, 134

\bibitem[{{Schmutz} {et~al.}(1989){Schmutz}, {Hamann}, \&
  {Wessolowski}}]{1989A&A...210..236S}
{Schmutz}, W., {Hamann}, W.-R., \& {Wessolowski}, U. 1989, \aap, 210, 236

\bibitem[{{Tylenda} {et~al.}(2003){Tylenda}, {Si{\'o}dmiak}, {G{\'o}rny},
  {Corradi}, \& {Schwarz}}]{2003A&A...405..627T}
{Tylenda}, R., {Si{\'o}dmiak}, N., {G{\'o}rny}, S.~K., {Corradi}, R.~L.~M., \&
  {Schwarz}, H.~E. 2003, \aap, 405, 627

\bibitem[{{van Hoof} {et~al.}(2008){van Hoof}, {Hajduk}, {Zijlstra}, {Herwig},
  {van de Steene}, {Kimeswenger}, \& {Evans}}]{2008ASPC..391..155V}
{van Hoof}, P.~A.~M., {Hajduk}, M., {Zijlstra}, A.~A., {et~al.} 2008, in
  Astronomical Society of the Pacific Conference Series, Vol. 391,
  Hydrogen-Deficient Stars, ed. A.~{Werner} \& T.~{Rauch}, 155

\bibitem[{{Weidmann} \& {Gamen}(2011)}]{2011A&A...526A...6W}
{Weidmann}, W.~A. \& {Gamen}, R. 2011, \aap, 526, A6

\bibitem[{{Wesson} {et~al.}(2005){Wesson}, {Liu}, \&
  {Barlow}}]{2005MNRAS.362..424W}
{Wesson}, R., {Liu}, X.-W., \& {Barlow}, M.~J. 2005, \mnras, 362, 424

\bibitem[{{Zijlstra} {et~al.}(2008){Zijlstra}, {van Hoof}, \&
  {Perley}}]{2008ApJ...681.1296Z}
{Zijlstra}, A.~A., {van Hoof}, P.~A.~M., \& {Perley}, R.~A. 2008, \apj, 681,
  1296

\end{thebibliography}

\end{document}